\documentclass{elsart}
\usepackage{epsfig}

\begin{document}

\begin{frontmatter}

\title{Random Matrix Theory and Chiral Logarithms}

\author[KL]{M.E.~Berbenni-Bitsch},
\author[R]{M.~G\"ockeler},
\author[R]{H.~Hehl},
\author[KL]{S.~Meyer},
\author[R]{P.E.L.~Rakow},
\author[R]{A.~Sch\"afer}, and
\author[M]{T.~Wettig}

\date{29 January 1999}

\address[KL]{Fachbereich Physik -- Theoretische Physik, Universit\"at
  Kaiserslautern, D-67663 Kaiserslautern, Germany}
\address[R]{Institut f\"ur Theoretische Physik, Universit\"at
  Regensburg, D-93040 Regensburg, Germany}
\address[M]{Institut f\"ur Theoretische Physik, Technische
  Universit\"at M\"unchen, D-85747 Garching, Germany}

\begin{abstract}
  Recently, the contributions of chiral logarithms predicted by
  quenched chiral perturbation theory have been extracted from lattice
  calculations of hadron masses. We argue that a detailed comparison
  of random matrix theory and lattice calculations allows for a
  precise determination of such corrections. We estimate the relative
  size of the $m\log(m)$, $m$, and $m^2$ corrections to the chiral
  condensate for quenched SU(2).  \\[2ex]

\end{abstract}

\end{frontmatter}

The identification of logarithmic corrections in the quark mass
predicted by quenched chiral perturbation theory \cite{Golt,Sharp} in
lattice gauge results is a long standing problem. It seems that the
latest numerical results \cite{Bard,Schier,Lat1,Lat2} on hadron masses
in quenched lattice simulations allow for an approximate determination
of these $\log(m)$ contributions. The determination of these
logarithms is an important test of chiral perturbation theory which in
turn plays a central role for the connection of low-energy hadron
theory on one side and perturbative and lattice QCD on the other.

In a completely independent development, it has been shown by several
authors that chiral random matrix theory (chRMT) is able to reproduce
quantitatively microscopic spectral properties of the Dirac operator
obtained from QCD lattice data (see the reviews~\cite{r1,r2} and
Refs.~\cite{r3,r4,r5,Ma}). Moreover, the limit up to which the
microscopic spectral correlations can be described by random matrix
theory (the analogue of the ``Thouless energy'') was analyzed
theoretically in \cite{Zahed98,Jac98} and identified for quenched
SU(2) lattice calculations in \cite{Ber2}.

The following analysis uses the scalar susceptibilities, so we first
give their definitions. The disconnected susceptibility is defined on
the lattice by
\begin{equation}
  \chi^{\rm disc}_{\rm lattice}=\frac{1}{N}\left\langle\sum_{k,l=1}^N
    \frac{1}{({\rm i}\lambda_k+m)({\rm i}\lambda_l+m)}\right\rangle -
  \frac{1}{N} \left\langle\sum_{k=1}^N\frac{1}{{\rm
        i}\lambda_k+m}\right\rangle^2 \;,
\label{e4}
\end{equation}
where $N=L^4$ denotes the number of lattice points and the $\lambda_k$
are the Dirac eigenvalues. After rescaling the susceptibility by
$N\Sigma^2$ ($\Sigma = $ absolute value of the chiral condensate for
infinite volume and vanishing mass) chRMT predicts
\begin{eqnarray}
  \frac{\chi^{\rm disc}_{\rm RMT}}{N\Sigma^2} &=& 4u^2\int_0^1\d
  s\:s^2K_0(2su) \int_0^1\d t\:I_0(2stu)\bigg\{s(1-t^2) \nonumber \\ 
  &&\qquad{}+4K_0(2u)[I_0(2su)+tI_0(2stu)] -
  8stI_0(2stu)K_0(2su)\bigg\} \nonumber \\ 
  &&\qquad{}-4u^2K_0^2(2u)\left[\int_0^1\d s\:I_0(2su)\right]^2 \\
  &=& 1-K_0(2u)I_0(2u)+[K_0(2u)-2uK_1(2u)]\int_0^1\d t\:I_0(2tu)
      \nonumber \\
  &&\qquad{}-\Bigg\{2uK_0(2u)\int_0^1\d t\:I_0(2tu) \nonumber \\
  &&\qquad{}-2u[K_0(2u)I_0(2u)+K_1(2u)I_1(2u)]\Bigg\}^2 \;,
\end{eqnarray}
where the rescaled mass parameter $u$ is given by $u=m\Sigma
L^4$. (For details we refer to \cite{Ber2}.)

\begin{figure}
  \centerline{\epsfig{figure=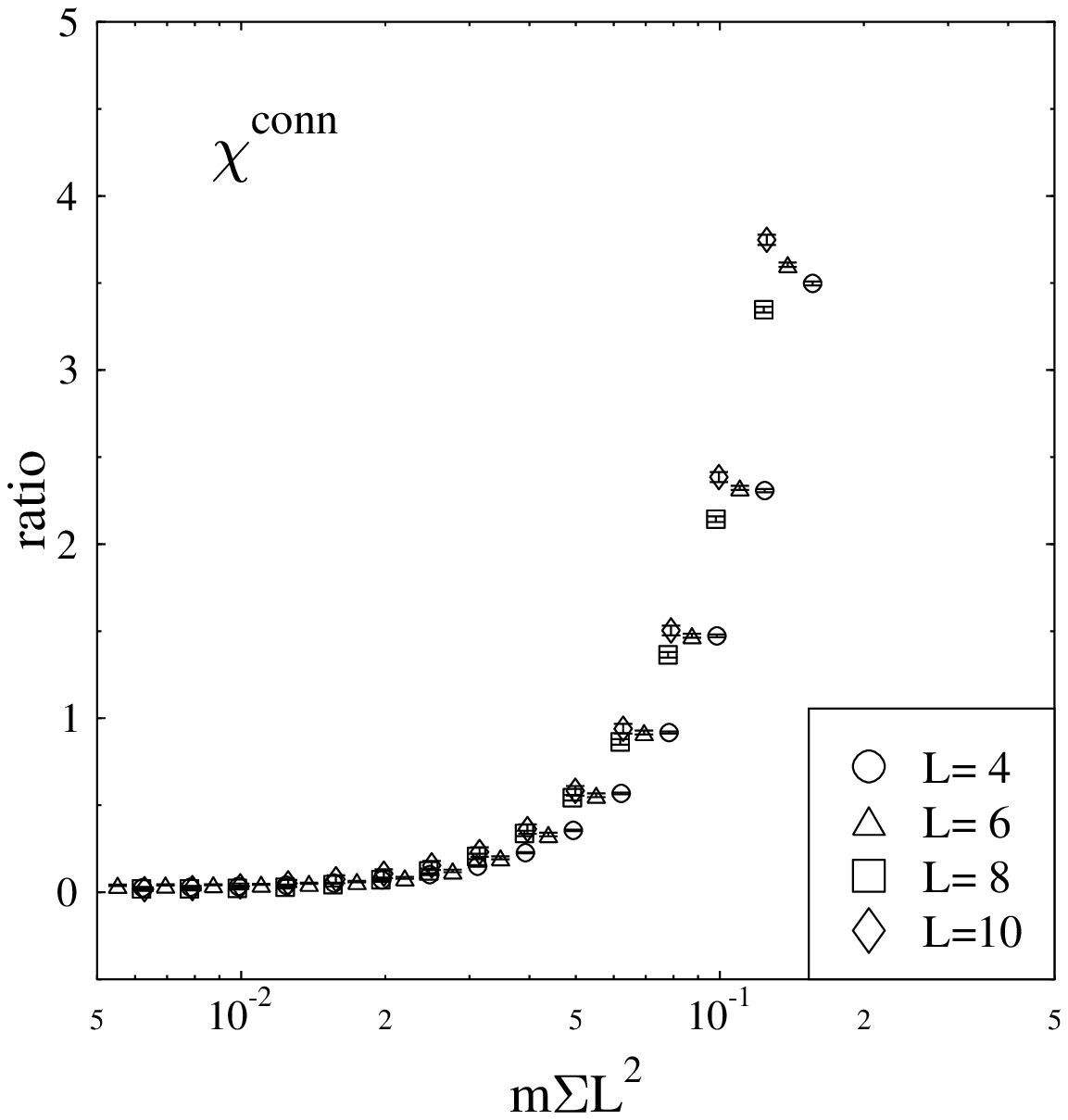,width=80mm}
    \epsfig{figure=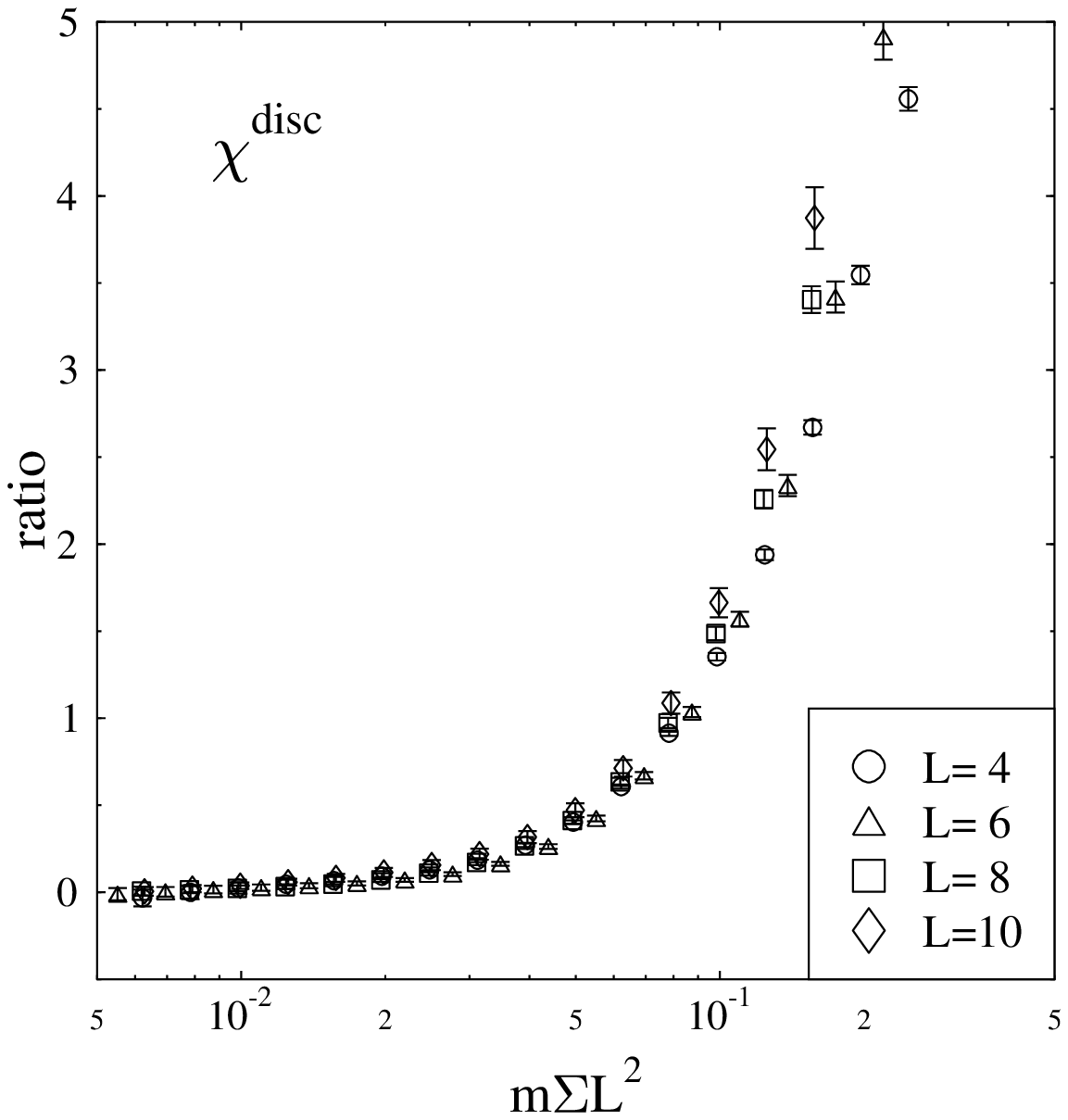,width=80mm}}
  \caption{The ratio of Eq.~(\protect\ref{ratio}) for the scaled
    susceptibilities plotted versus $m\Sigma L^2$ (in lattice units)
    for $\beta=2.0$ and four different lattice sizes, $N=4^4$, $6^4$,
    $8^4$, and $10^4$.}
  \label{scalchidSU2}
\end{figure}

We shall also use the connected susceptibility which is defined on the
lattice by
\begin{equation}
  \chi^{\rm conn}_{\rm lattice} =
  -\frac1N\left\langle\sum_{k=1}^N\frac1{({\rm
        i}\lambda_k+m)^2}\right\rangle\;.
\end{equation}
The chRMT result reads
\begin{equation}
  \frac{\chi^{\rm conn}_{\rm RMT}}{N\Sigma^2} = 4uK_1(2u)\int_0^1\d
  s\:(1-s)I_0(2su)\;.
\end{equation}
Fig.~\ref{scalchidSU2} presents the deviation of the (parameter-free)
random matrix prediction from the lattice result, more precisely the
ratio
\begin{equation}
  {\bf ratio}=\left( \chi_{\rm lattice}-\chi_{\rm RMT} \right) /
  \left( \chi_{\rm RMT} \right)\;,
  \label{ratio}
\end{equation}
where $\chi$ can either be the disconnected (only this choice was
investigated in \cite{Ber2}) or the connected susceptibility.

The motivation for investigating ${\bf ratio}$ rather than $\chi_{\rm
  lattice}$ itself is that in Eq.~(\ref{ratio}) finite size
corrections cancel to a remarkable degree, allowing us to use data
from smaller $m$ values. We have seen in Fig.~2 of \cite{lat97} that
the knowledge of finite size effects which we gain from RMT allows us
to find the thermodynamic limit of the chiral condensate from
extremely small lattices. This can also be formulated in the following
way: for a given value of ${\bf ratio}$ in Fig.~\ref{scalchidSU2}, the
finite size corrections for all four lattice sizes are expected to be
similar, as the corresponding values of $m_\pi^2 L^2 \propto mL^2$ are
very close, which is why we have plotted ${\bf ratio}$ against
$m\Sigma L^2$ in Fig.~\ref{scalchidSU2}.

What do we expect beyond the Thouless energy? Then, the lattice is
large enough so that the valence pion, which is the lightest particle,
fits on the lattice. Naturally, all other particles also fit on the
lattice, and therefore we expect that the chiral condensate and the
two susceptibilities will rapidly approach their thermodynamic limit.

For a finite lattice and a non-vanishing mass, the chiral condensate
is given by
\begin{equation}
 \sigma_{\rm lattice}(m) = \frac{1}{N} \left\langle
 \sum_{k=1}^N \frac{1}{{\rm i}\lambda_k+m} \right\rangle \;.
\end{equation}
In the quenched theory, the connected susceptibility is given simply
by
\begin{equation}
 \chi^{\rm conn}(m) = \frac{\partial}{\partial m} \sigma(m) \;,
\end{equation}
so we can find the infinite-volume behavior of $\chi^{\rm conn}$ from
that of $\sigma$. We expect from chiral perturbation theory
\cite{chiralpert} that the chiral condensate has the form
\begin{equation}
 \sigma(m) = \Sigma \left[ 1 - A m \log (m) + B m
  + \frac{1}{2} C m^2 + \cdots \right] \;.
 \label{sigtherm}
\end{equation}

Eq.~(\ref{sigtherm}) requires several comments. In the continuum,
quenched chiral perturbation theory predicts a leading term
proportional to $\frac{\langle\nu^2\rangle} {L^4} \log(m)$, where
$\langle\nu^2\rangle /L^4$ is the topological susceptibility
\cite[Sec.~7]{chiralpert}. We argue that this leading term should be
absent in our case. For finite lattice spacing the
Atiyah--Singer-index theorem does not apply for staggered fermions.
Therefore the role of topology has to be interpreted with care. We
have seen in \cite{r5} that the small Dirac eigenvalues are well
described by random matrix results for $\nu=0$.  This means that the
quasi-zero modes related to topology are shifted to such large values
that they are not visible. (This is presumably due to discretization
errors proportional to $a^2$, with $a$ the physical lattice spacing.)
Thus the violation of axial symmetry which generates the logarithmic
term in the quenched case is dominated by the explicit quark masses,
which motivates Eq.~(\ref{sigtherm}). It would be very interesting to
study the $\nu\neq 0$ sector for which we expect a leading $\log(m)$
term, which might require, however, very small $a$ and a large number
of lattice points.

Eq.~(\ref{sigtherm}) implies that in the thermodynamic limit 
\begin{equation}
 \chi^{\rm conn}_{\rm lattice} =
 \Sigma \left[ -A \log (m) -A + B + C m + \cdots \right]\;.
 \label{chilatt}
\end{equation}
On the other hand, the large-volume limit of the RMT susceptibility is
\begin{equation}
 \chi^{\rm conn}_{\rm RMT} \rightarrow \frac{1}{4 m^2 L^4} \;. 
\end{equation}
Putting the two expressions together, we find that
\begin{equation}
  {\rm {\bf ratio }} \rightarrow (m \Sigma L^2 )^2 \frac{4}{\Sigma}
  \left[ -A \log (m) -A + B + C m + \cdots \right] -1\;.
 \label{thermo}
\end{equation}
Strictly speaking, the $-1$ ought to be neglected in comparison with
the first term as $L\to\infty$. However, we should be prepared to
observe some sub-leading corrections in the data taken on finite
lattices.

We have confronted Eq.~(\ref{thermo}) with lattice Monte Carlo data
for two values of the coupling strength $\beta$, $\beta=2.0$ and
$\beta=2.2$. The lattice sizes and numbers of configurations are given
in Table~\ref{numcon}.
\begin{table}[hb]
\caption{Lattice sizes and numbers of configurations for the lattice
  data.}
\label{numcon}
\begin{center}
\begin{tabular}{lcccc}\\
\hline\hline
\multicolumn{5}{c}{$\beta=2.0$}\\
\hline
 L             &   4   &   6   &   8   &  10  \\
 \# of configs & 49978 & 24976 & 14290 & 4404 \\
\hline\hline
\multicolumn{5}{c}{$\beta=2.2$}\\
\hline
 L             &   6   &   8   &   10 &  12  \\
 \# of configs & 22292 & 13975 & 2950 & 1388 \\
\hline\hline
\end{tabular}
\end{center}
\end{table}

To check Eq.~(\ref{thermo}) we did the following for both values of
$\beta$:\\ 
We chose different values for ${\bf ratio}=b_i$ and determined the
values of $m\Sigma L^2$ for which they were reached for our different
lattice sizes.  Let us denote these numbers by $Y(L,b_i)$.
Eq.~(\ref{thermo}) implies that
\begin{equation}
 \frac{1}{Y(L,b_i)^2} =
 r(b_i)\left[ -\log (m) +\frac BA -1 + \frac CA m + \cdots \right]
 \label{yfit}
\end{equation}
where $r(b)$ will be proportional to $1/b$ as $b\to\infty$. Since we
do not reach too large values of $b$, we used the ansatz
\begin{equation}
 \frac{1}{Y(L,b_i)^2} =
 \frac q{b_i+s}\left[ -\log (m) +\frac BA -1 + \frac CA m + \cdots
 \right]
 \label{ansatz}
\end{equation}
to fit our data. In Eq.~(\ref{ansatz}) not only $Y^{-2}$ has
statistical errors, but also $m$. In our $\chi^2$ fit, however, only
the errors of $Y^{-2}$ are taken into account.

Obviously, the values $Y(L,b_i)$ for the same lattice size $L$ are
highly correlated. It is, however, unclear how to calculate the
correlations of these quantities, which are related to the original
lattice results only in a rather implicit manner.  Moreover,
correlated fits tend to have problems \cite{CMich}.  Therefore we
decided to ignore correlations completely, although this will lead to
an underestimation of the errors on the fit parameters.

For the thermodynamic limit of the disconnected susceptibility we
assume the same form as Eq.~(\ref{chilatt}). In RMT, the large-volume
limit is given by
\begin{equation}
  \chi^{\rm disc}_{\rm RMT} \to \frac 1 {8m^2L^4}
\end{equation}
so that the ansatz of Eq.~(\ref{ansatz}) applies as well.

\begin{figure}
  \centerline{\epsfig{figure=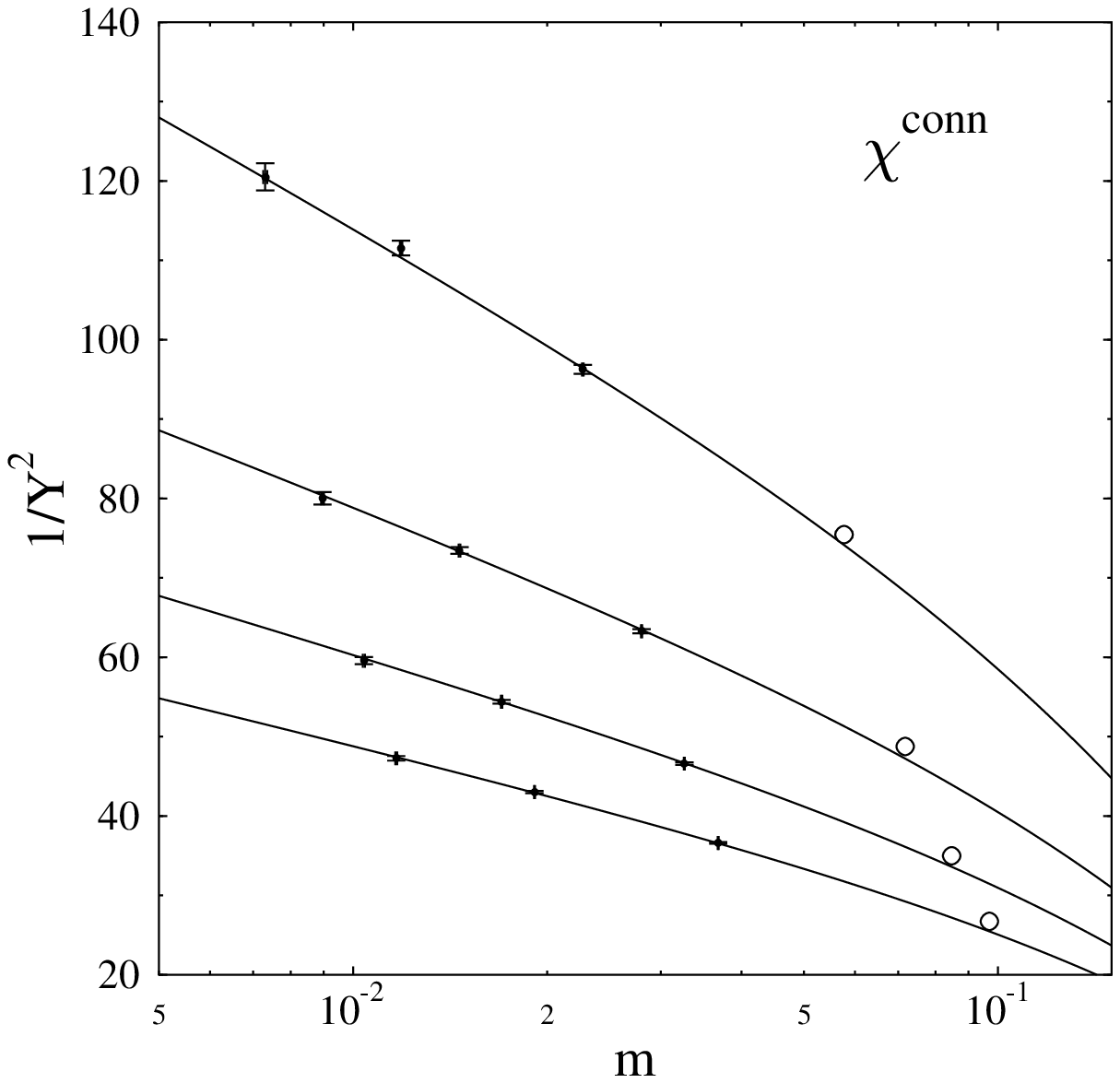,width=80mm}
    \epsfig{figure=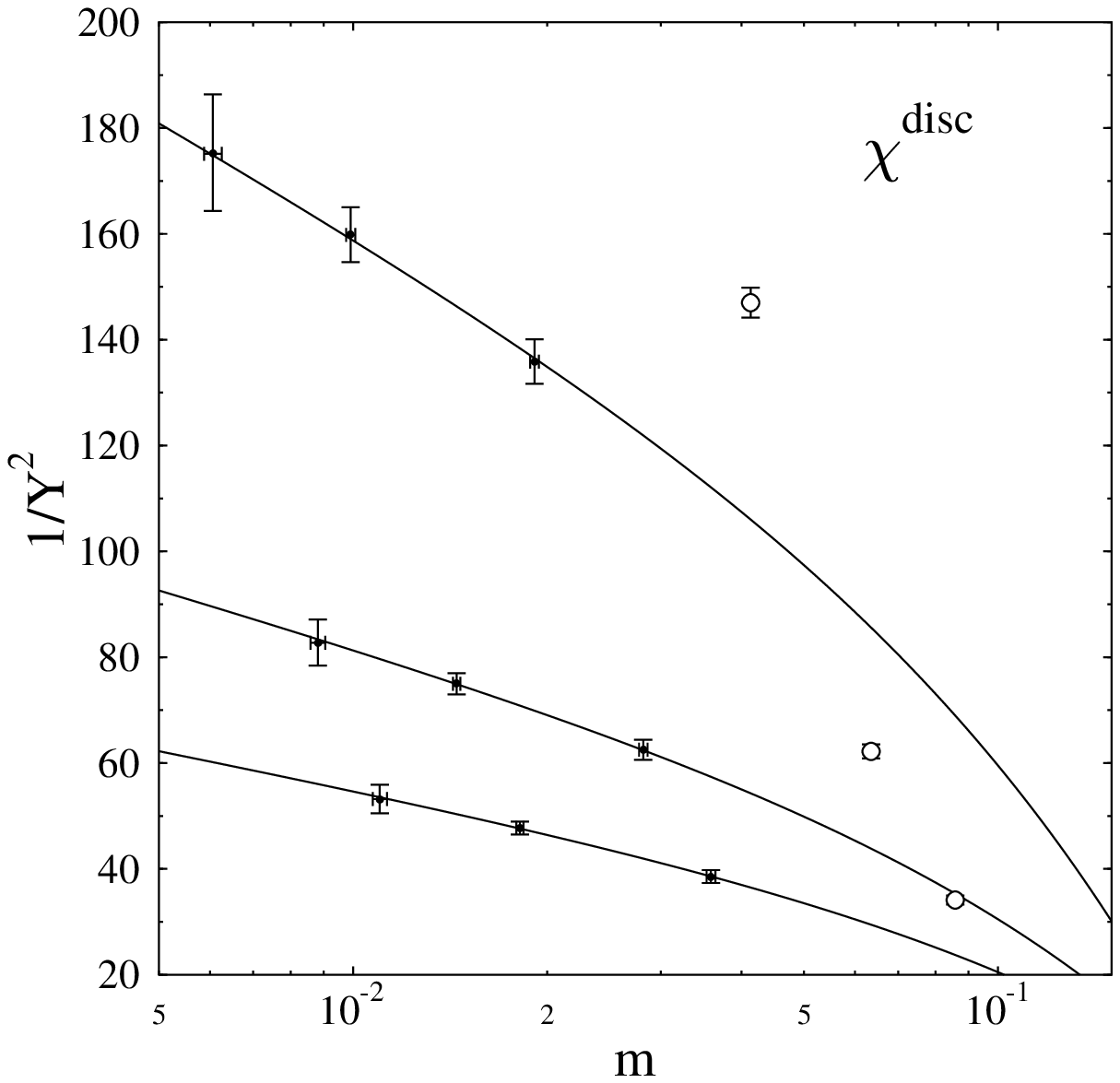,width=80mm}}
  \caption{The value of $Y=m\Sigma L^2$ at $\beta = 2.0$ for which
    $\mathbf{ratio}=b$ for various values of $b$ as a function of $m$.
    Larger values of $b$ belong to smaller values of $1/Y^2$. The
    rightmost filled dots correspond to $L=6$, the leftmost to $L=10$,
    whereas the open dots represent data for $L=4$, which were not
    used in the fit. All quantities are measured in lattice units.}
  \label{pl-2p0}
\end{figure}

\begin{figure}
  \centerline{\epsfig{figure=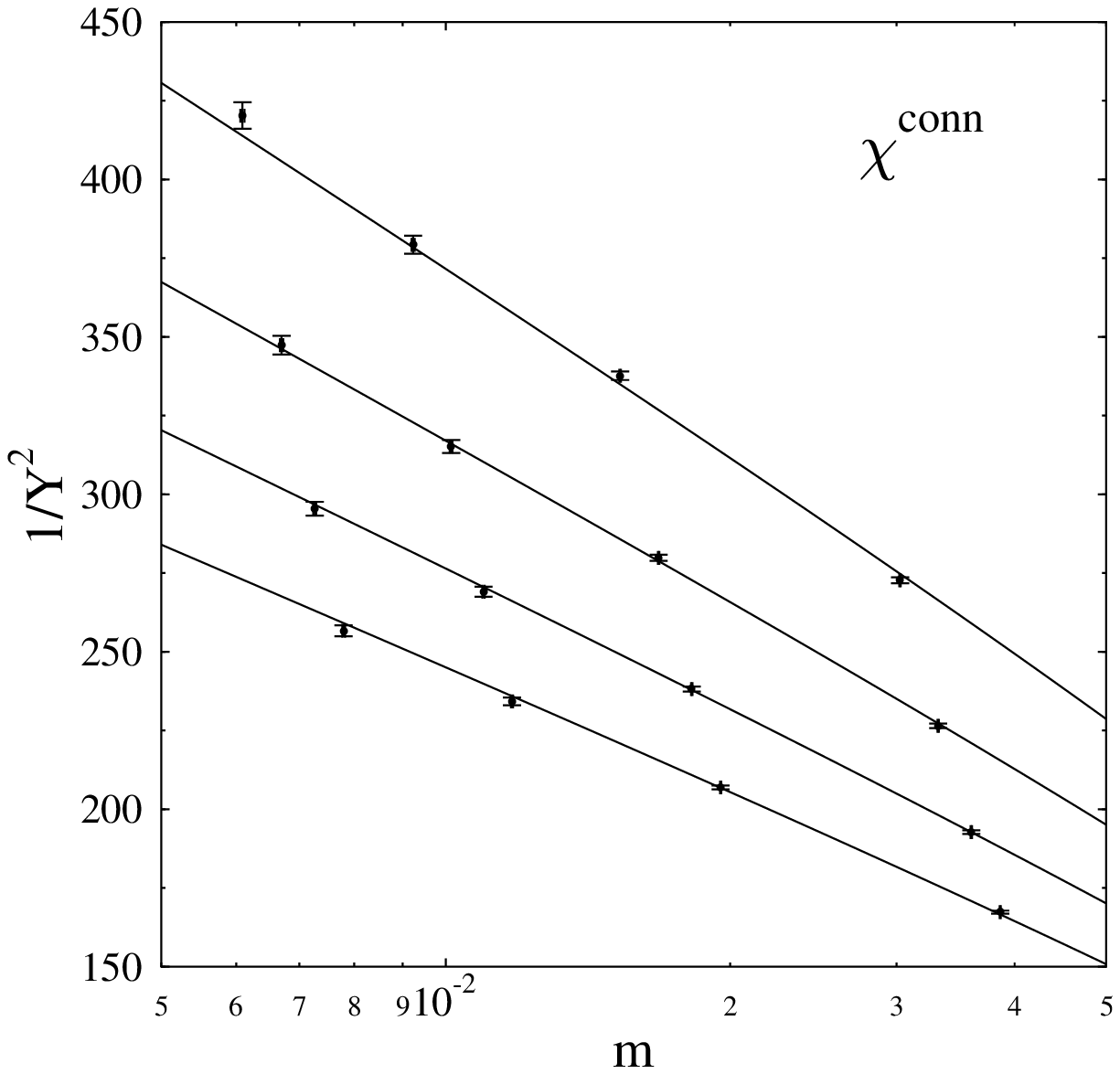,width=80mm}
    \epsfig{figure=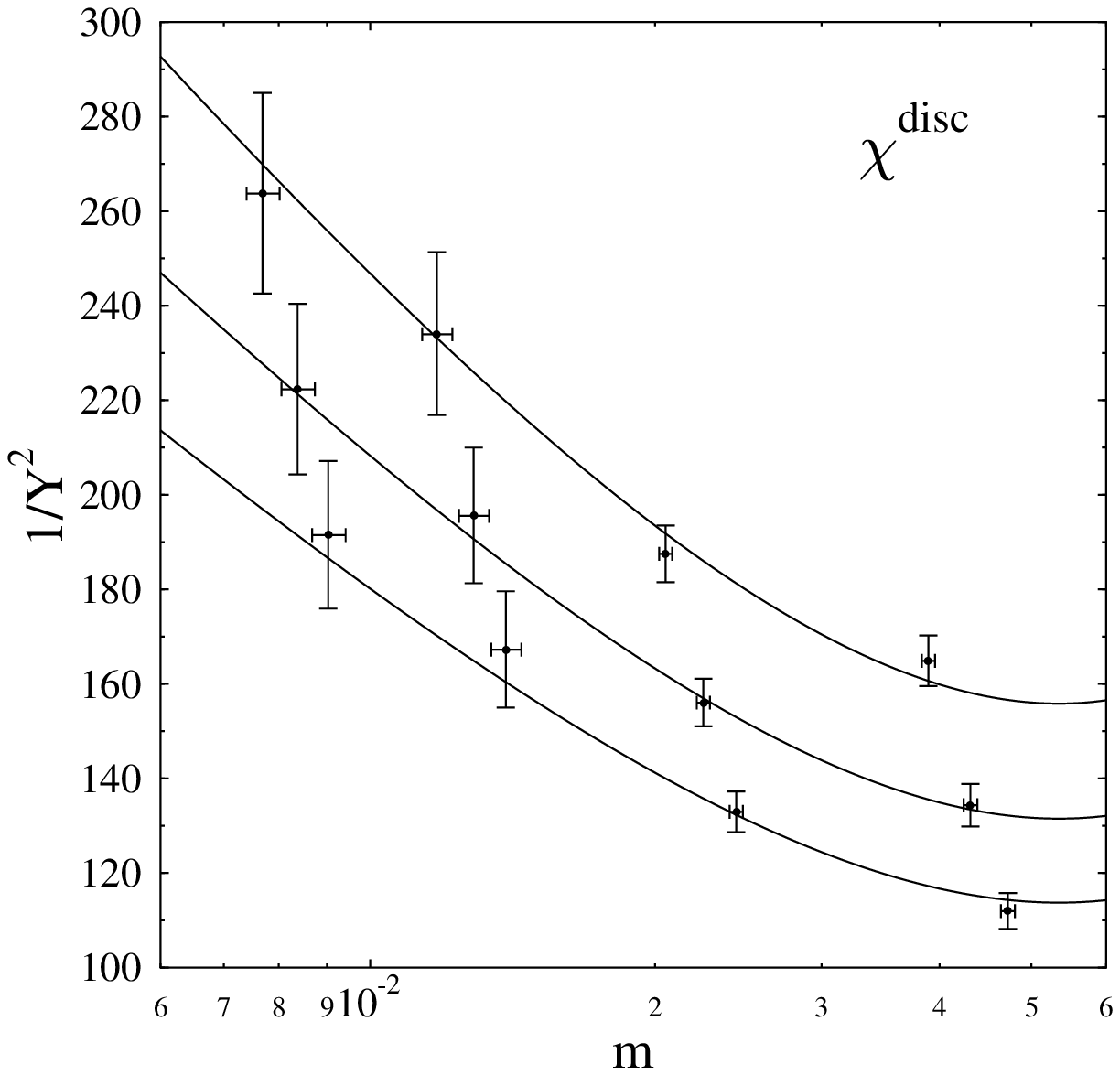,width=80mm}}
  \caption{Same as Fig.~\protect\ref{pl-2p0} but for $\beta=2.2$ and
    with the dots, from left to right, corresponding to $L=12$, $10$,
    $8$, and $6$, respectively.}
  \label{pl-2p2}
\end{figure}

In Figs.~\ref{pl-2p0} and \ref{pl-2p2} we plot $Y^{-2}$ versus $m$
together with the fits for $\beta = 2.0$ and $2.2$, respectively. In
the case of the connected susceptibility we used $b_i = 2.0, 3.0, 4.0,
5.0$ ($\beta=2.0$) and $b_i = 5.0, 6.0, 7.0, 8.0$ ($\beta=2.2$) and
obtained the results of Table~\ref{tableconn}.
\begin{table}[b]
\caption{Fit parameters for the connected susceptibility.}
\label{tableconn}
\begin{center}
\begin{tabular}{cr@{${}\pm{}$}lr@{${}\pm{}$}lr@{}r@{${}\pm{}$}r@{}l
    r@{${}\pm{}$}lc}\\
\hline
$\beta$ & \multicolumn{2}{c}{$B/A$} & \multicolumn{2}{c}{$C/A$} &
\multicolumn{4}{c}{$q$} & \multicolumn{2}{c}{$s$} &
$\chi^2/\mathrm{dof}$ \\
\hline
2.0 & 2.29 & 0.63 & $-5.97$ & 5.17 &  43&.9&  4&.5 & 0.25 & 0.02
   & 0.50 \\
2.2 & 0.86 & 0.18 & $-2.46$ & 1.86 & 486&  & 19&   & 0.81 & 0.05
   & 1.75 \\
\hline
\end{tabular}
\end{center}
\end{table}
For the disconnected susceptibility we used $b_i = 1.0, 2.0, 3.0$
($\beta=2.0$) and $b_i = 6.0, 7.0, 8.0$ ($\beta=2.2$) and found the
values given in Table~\ref{tabledisc}.
\begin{table}[b]
\caption{Fit parameters for the disconnected susceptibility.}
\label{tabledisc}
\begin{center}
\begin{tabular}{cr@{.}l@{${}\pm{}$}lr@{}r@{${}\pm{}$}r@{}l
    r@{${}\pm{}$}rr@{${}\pm{}$}lc}\\
\hline
$\beta$ & \multicolumn{3}{c}{$B/A$} & \multicolumn{4}{c}{$C/A$} &
\multicolumn{2}{c}{$q$} & \multicolumn{2}{c}{$s$} &
$\chi^2/\mathrm{dof}$ \\
\hline
2.0 &    1&9  & 3.1  & $-12$&   & 32&   &  31 &  16 &   0.05
   & 0.05 & 0.02 \\
2.2 & $-1$&45 & 0.48 &   18 &.7 &  4&.1 & 569 & 127 & $-0.60$
   & 0.61 & 0.28 \\
\hline
\end{tabular}
\end{center}
\end{table}

The main message of Figs.~\ref{pl-2p0} and \ref{pl-2p2} is that
without any doubt the data are not fitted by horizontal lines.  This
demonstrates the presence of additional contributions in the quark
mass. The approximate linearity of the curves for small $m$ shows that
the logarithmic contribution is the dominant one. For the connected
susceptibility, the data are well fitted by the ansatz
(\ref{sigtherm}), i.e., with only the three leading corrections. For
the disconnected susceptibility, our statistical precision does not
allow for a precise determination of the ratios $B/A$ and $C/A$. For
very small lattices ($4^4$ in Fig.~\ref{pl-2p0}) finite size effects
seem to spoil our analysis.

It is clear from Figs.~\ref{pl-2p0} and \ref{pl-2p2} that one would
really like to have numerical simulations with substantially larger
statistics and larger lattices. As the applicability of RMT to the
description of the low-energy Dirac spectrum is by now well
established we can limit ourselves in the future to the calculation of
just the lowest eigenvalues instead of the complete spectrum. This
should allow us to gain the necessary statistics.

To conclude, let us remark that the aim of this paper is primarily to
draw attention to this new method to extract chiral logarithms and
other corrections in the quark mass, and to stimulate the discussion
of their interpretation. The obvious next step is to analyze the
susceptibilities within the framework of quenched chiral perturbation
theory.

\begin{ack}
  It is a pleasure to thank I.~Zahed for an early discussion and
  M.~Golterman, N.~Kaiser, and J.J.M.~Verbaarschot for helpful
  comments. This work was supported in part by DFG. SM, AS, and TW
  thank the MPI f\"ur Kernphysik, Heidelberg, for hospitality and
  support. The numerical simulations were performed on a CRAY T90 at
  the Forschungszentrum J\"ulich, on a CRAY T3E at the HLRS Stuttgart,
  and on a Cray T90 at the Leibniz-Rechenzentrum M\"unchen.
\end{ack}

\end{document}